\documentclass[lettersize,onecolumn]{IEEEtran}
\usepackage{amsmath,amsfonts}
\usepackage{algorithmic}
\usepackage{algorithm}
\usepackage{array}
\usepackage{textcomp}
\usepackage{stfloats}
\usepackage{url}
\usepackage{verbatim}
\usepackage{subcaption}
\usepackage{graphicx}
\usepackage{cite}
\usepackage{ifthen}
\usepackage{color}
\usepackage{amssymb}
\usepackage{caption}
\usepackage{adjustbox}
\usepackage{booktabs}
\usepackage{rotating}
\usepackage{mathrsfs}

\begin{document}

\title{Optimizing Sequencing Coverage Depth in DNA Storage: Insights From DNA Storage Data}

\author{Ruiying Cao, Xin Chen}



\maketitle

\begin{abstract}
DNA storage is now being considered as a new archival storage method for its durability and high information density, but still facing some challenges like high costs and low throughput. By reducing sequencing sample size for decoding digital data, minimizing DNA coverage depth helps lower both costs and system latency. Previous studies have mainly focused on minimizing coverage depth in uniform distribution channels under theoretical assumptions. In contrast, our work uses real DNA storage experimental data to extend this problem to log-normal distribution channels, a conclusion derived from our PCR and sequencing data analysis. In this framework, we investigate both noiseless and noisy channels. We first demonstrate a detailed positive correlation between MDS code rate and the expected minimum sequencing coverage depth. Moreover, we observe that the probability of successfully decoding all information in a single sequencing run decreases and then increases as code rate rises, when the sample size is optimized for complete decoding. Then we extend the lower bounds of the DNA coverage depth from uniform to log-normal noisy channels. The findings of this study provide valuable insights for the efficient execution of DNA storage experiments. 
\end{abstract}


\section{Introduction}
With the rapid growth of global data and advancements in information technology, traditional storage media such as HDDs and SSDs may no longer meet future data storage demands \cite{ref1,ref2}. DNA storage, with its high density and durability, has emerged as a promising solution to address this challenge. However, current DNA storage technologies face significant obstacles, including low throughput and high costs. Reducing the sequencing sample size required to ensure a high probability of decoding all information is the main goal of coverage depth problem, which could provide valuable insights for reducing latency and associated costs.

Several studies have addressed this issue in DNA storage channels follow uniform distribution by adjusting outer error-correcting codes. These approaches are often based on classical problems such as the coupon collector's problem\cite{ref3}, the urn problem\cite{ref4}, and others. \cite{ref5} first introduced the concept of sequencing coverage depth, investigating the minimum coverage depth required for successful message decoding under non-random access settings. They also explored the singleton coverage depth problem and proposed a new coding scheme to reduce the sample size needed to retrieve a single strand. \cite{ref6} extended the work of \cite{ref5} to more practical application scenarios, particularly focusing on retrieving $k$ (where $k=1$ or $2$) out of $n$ files. This study also examined the structural properties of various coding schemes and their impact on random access expectations and probability distributions. \cite{ref7} provided a novel perspective on random access efficiency by exploring coding matrix design and performance optimization. Their work introduced the concept of recovery balanced codes and corresponding conditions, offering theoretical guidance for designing more efficient DNA storage systems. \cite{ref8} proposed the first model to analyze sequencing coverage depth in combinatorial coding, calculating the probability of error-free reconstruction and investigating theoretical bounds to evaluate decoding probabilities. \cite{ref9} introduced a new geometric structure, namely the balanced quasi-arc, to study the random access problem. By analyzing this geometric structure, the study aimed to identify encoding schemes that could effectively reduce the expected number of samples required for random access.

These studies collectively address the problem of minimizing sequencing coverage depth in uniformly distributed channels, offering various theoretical models, coding schemes, and analytical methods. Their contributions provide a foundation for optimizing sequencing efficiency and cost-effectiveness in DNA storage systems. However, the processes of DNA synthesis and amplification exhibit a degree of randomness, which results in a non-uniform channel probability distribution. Therefore, in this study, we investigate the problem of minimizing sequencing coverage depth in a real-world channel based on PCR (Polymerase Chain Reaction) and sequencing data, under the non-random access setting.

This paper is organized as follows: Section \uppercase\expandafter{\romannumeral2} provides an overview of related work and offers a detailed description of the problem addressed in this study. Section \uppercase\expandafter{\romannumeral3} analyzes PCR and sequencing data to establish the foundation for research on minimizing sequencing coverage depth in channels where the probability distribution follows a log-normal distribution. Section \uppercase\expandafter{\romannumeral4} investigates the expected value of minimum sequencing coverage depth in the noiseless channel and two lower bounds in the noisy channel, under the non-random access setting.

\section{Problem Statement, Related Work}
\subsection{Problem Statement}
The problem we study is built on the following DNA storage model (see Fig. 1).
\begin{figure*}
  \includegraphics[width=\linewidth]{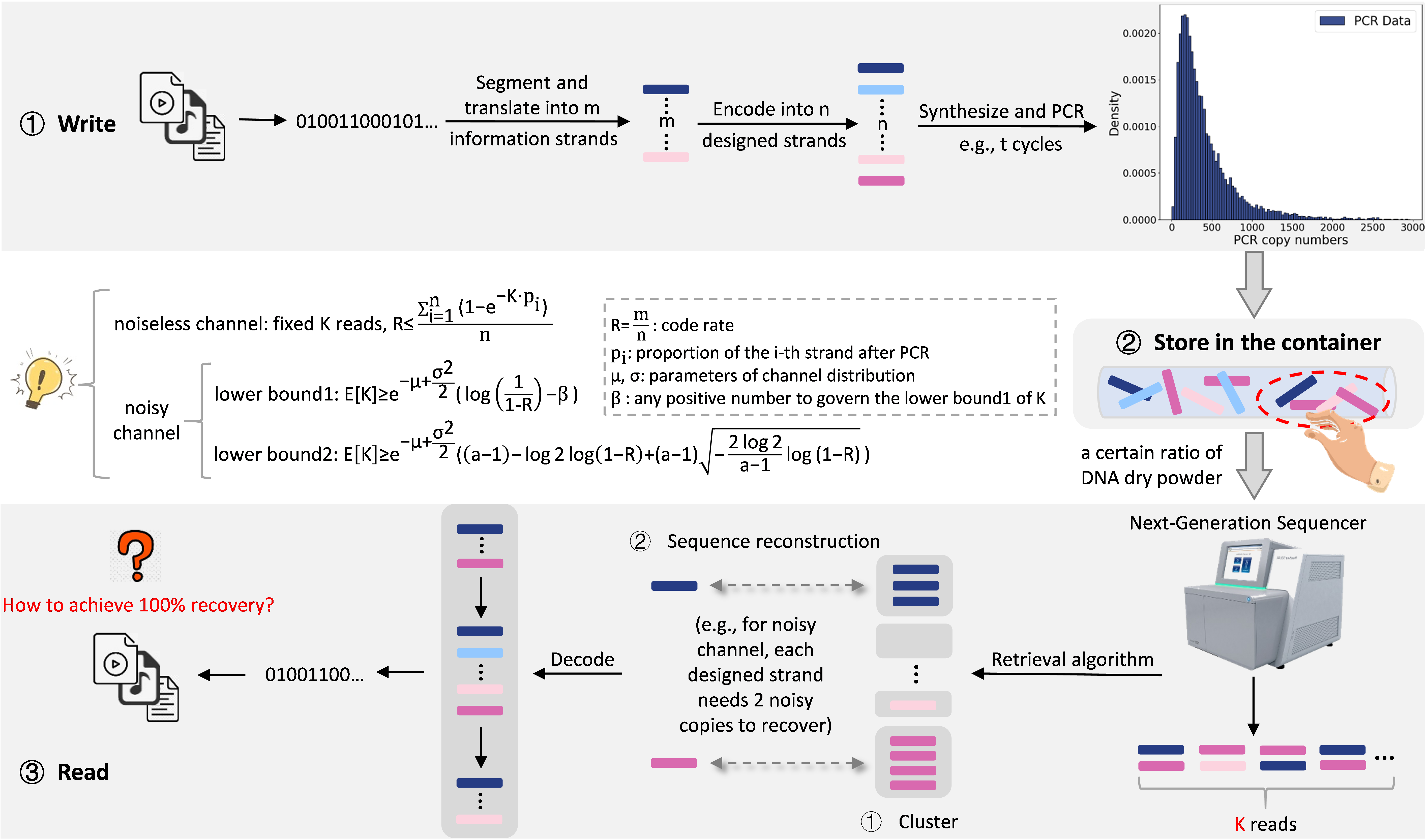}
  \caption{Framework of this paper.}
  \label{fig:rotated}
\end{figure*}

For alphabet $\varSigma=\{A, C, G, T\}$, each DNA strand of length $l$ can be represented as a vector of length $l$ over the alphabet $\varSigma$. The set of all such vectors of length $l$ over $\varSigma$ is denoted by $\varSigma^{l}$. Let $\varSigma^{*}\triangleq \bigcup_{l=0}^{\infty} \varSigma^{l}$, and $[1,n]=1, 2, \dots, n$.

\textbf{Write and encode}. Assume that the digital data to be stored is first segmented and translated into $m$ information strands of length $l$, denoted as $X=(x_1, x_2, \dots , x_m) \in (\varSigma^{l})^m$. After encoding with an $[n,m]$ MDS code\cite{ref10}, the $m$ information strands are transformed into $n$ designed strands of length $l$, represented as $D=(d_1, d_2,\dots,d_n)\in (\varSigma^{l})^n$. Next, each designed strand in $D$ is synthesized into $c$ copies. Due to the inherent randomness in the synthesis process, the actual copy numbers of each designed strand $c_i$ is not the same \cite{ref11}, where $i \in [1,n]$, and each synthesis strand is a noisy version of its corresponding designed strand. Then all synthetic strands will be stored unordered in a container until reading them out.

\textbf{Read and decode}. In this step, we will extract a specific volume or weight of DNA dry powder in the container and performing sequencing using next-generation sequencer (NGS). Let the multiple set $S_K=\{\{s_1,s_2,\dots,s_K\}\}$ represents the $K$ reads obtained from sequencing, where each $s_i \in \varSigma^{*}$ is a noisy copy of a particular designed strand $d_j$. The number of noisy copies of the $i$-th designed strand $d_i$ in $S_K$ depends on the probability distribution $\boldsymbol{p}_t=\left(p_1^{(t)},p_2^{(t)},\dots,p_n^{(t)}\right)$, where $p_i^{(t)}$ represents the probability of sampling a read of $d_i$ in each round, i.e., the proportion of $d_i$ in the population after $t$ cycles of PCR amplification. During the decoding process, we first apply a clustering algorithm to partition the reads in $S_K$ into clusters based on their corresponding designed strands. Specifically, the $i$-th cluster contains all noisy copies of the $i$-th designed strand $d_i$ in $S_K$, denoted as $s_J=\{s_j|s_j \ \text{is a noisy copy of} \ d_j\}$. Then we use the retrieved designed strands to recover the rest information strands.\\
\textbf{Remark}. During the sequencing process, due to differences in sequencing technologies, the reads in $S_K$ can either be obtained sequentially, as in Nanopore sequencing, or all at once, as in Illumina sequencing. However, the research conclusions drawn from both approaches do not show significant differences \cite{ref5}. In this study, a sequential reading method is employed.

Notice that due to the inherent randomness in the synthesis, amplification, and sequencing processes of DNA storage \cite{ref11}, it is uncertain whether all the original information can be fully recovered from the $K$ reads obtained through sequencing. Thus, we focus on how to ensure the 100\% successful decoding of the original data with high probability, using as small sequencing sample size as possible in the real channel.

We formulate this coverage depth problem under non-random access setting as a variant of the classical coupon collector’s problem or the urn model. Specifically, we consider the scenario in which one identical ball is thrown in each round, and the probability of a ball falling into each urn varies. After throwing $K$ rounds, we are interested in the number of urns—among $n$ indistinguishable urns except for their labels—that contain at least $a$ balls. To this end, we model the sample size as a function of the channel probability distribution, the MDS code, and the retrieval algorithm. We proceed to give a detailed explanation of these three variables.
\begin{enumerate}
    \item Explanation of the channel probability distribution. The channel probability distribution is analogous to the probability distribution of a ball falling into each urn. Unlike \cite{ref5}, which assumes a uniform distribution of the number of each designed strands, i.e., the probability of a ball falling into each urn is equal, we consider that the randomness in synthesis and PCR amplification causes varying copy numbers of each designed strand, leading to unequal proportions in the population. We assume that the number of noisy copies of each designed strand observed in the sequencing reads follows a probability distribution $\boldsymbol{p}_t=\left(p_1^{(t)},p_2^{(t)},\dots,p_n^{(t)}\right)$, where $p_i^{(t)}$ denotes the probability of sampling a read corresponding to the $i$-th strand after $t$ cycles of PCR amplification. Notice that for simplicity, we consider the distribution $\boldsymbol{p}_t$ solely as a function of the DNA storage channel, without accounting for potential influences from strand design \cite{ref5}. Accordingly, we refer to $\boldsymbol{p}_t$ as the channel probability distribution in this paper.
    \item Explanation of the MDS code. The construction of the error-correcting code determines how many designed strands need to be retrieved to decode all digital data, i.e., how many urns are needed to contain at least $a$ balls after $K$ rounds? In our work, when an $[n,m]$ MDS code is used to encode the information strands, successful retrieval of any $m$ out of the $n$ encoded strands is sufficient to fully recover the original information. We denote the code rate of the $[n,m]$ MDS code by $R=\frac{m}{n}$.
    \item Explanation of the retrieval algorithm. The probability of successful retrieval of each designed strand depends on the number of its noisy copies in the sequencing pool, the use of inner codes during encoding, and the channel error rate \cite{ref5}. In this paper, we model the retrieval algorithm by introducing a positive integer parameter $a \geq 1$, assuming that each designed strand can be successfully retrieved if at least $a$ reads corresponding to it are available.
\end{enumerate}

Based on the assumptions and analysis outlined above, the main problems we investigate in this paper are defined below.\\
\textbf{Problem 1. (Simulation of DNA storage channel distribution.)} Given values of $n\ge m \ge 1$, $t \ge 1$, the synthesis amount of each designed strand $c_1,c_2,\dots,c_n \stackrel{i.i.d}{\sim} p(c)$, and the amplification efficiency of each designed strand $r_1,r_2,\dots , r_n$, we focus on the following questions: 
\begin{enumerate}
  \item The probability distribution of $\nu_i^{(t)}$, i.e., copy numbers of the $i$-th designed strand after $t$ cycles of PCR amplification.
  \item The probability distribution of $p_i^{(t)}=\frac{\nu_i^{(t)}}{\sum_{j=1}^n \nu_j^{(t)}}$, i.e., the proportion of the $i$-th designed strand in the population after $t$ cycles of PCR amplification.
\end{enumerate}
\textbf{Problem 2. (MDS coverage depth problem in the real channel.)} Given values of $n\ge m \ge 1$, $a\ge1$, we focus on the following questions:
\begin{enumerate}
  \item The expectation value $\mathbb{E}[K_a(n,m)]$.
  \item Given $t \ge 1$, the expectation value $\mathbb{E}[K_1^{\boldsymbol{p}_t}(n,m)]$.
  \item Given $t \ge 1$, the lower bounds of $K_a^{\boldsymbol{p}_t}(n,m)$.
\end{enumerate}
\subsection{Previous Work}
By directly mapping the coupon collector's problem, urn and dixie cup problem to the sequencing coverage depth problem, we obtain from \cite{ref12,ref13,ref14,ref15} that, 
\begin{equation*}
    \mathbb{E}[K_1(n=m,m)]=m \log m+\gamma m+\mathcal{O}(1),
\end{equation*}
where $\gamma \sim 0.577$ is the Euler–Mascheroni constant.
\begin{equation*}
    \mathbb{E}[K_1(n,m)]=n (H_n-H_{n-m}),
\end{equation*}
where $H_n$ is the $n$-th harmonic number.
\begin{align*}
    \mathbb{E}[K_a(n=m,m)]&=m \log m +m(a-1) \log \log m \\
    &+mC_a+o(m),
\end{align*}
where $C_a$ is a constant that corresponding to $a$.
\begin{equation}
\label{equ10}
    \mathbb{E}[K_a^{\boldsymbol{p}}(n,m)]=\sum_{q=0}^{m-1} \int_0^{\infty}[v^q] \prod_{i=1}^n (e_{t-1}(p_i r)+v(e^{p_i r}-e_{t-1}(p_i r)))e^{-r} dr,
\end{equation}
where $\boldsymbol{p}$ represents any general discrete distribution, $[v^q]Q(v)$ represents the coefficients of the $q$-th terms of the polynomial $Q(v)$, and $e_t (x)=\sum_{i=0}^t \frac{x^i}{i!}$.

Based on the models above, previous work has been carried out mainly in channels following uniform distribution. For example, \cite{ref3} proved that for any $\epsilon > 0$,
\begin{equation*}
    \log \left(\frac{1}{1-R}\right)+f_c(n,R) \le \mathbb{E}\left[\frac{K_a(n,m)}{n}\right] \le \left(\log\left(\frac{1}{1-R}\right)+a\log \log n +2\log(a+1)\right)\cdot(1+2\epsilon),
\end{equation*}
where $f_c(n,R)=\mathcal{O}(\frac{1}{n^2})$.

Equation (\ref{equ10}) does not provide a closed-form expression and is not straightforward to compute. Thus, we will build upon the MDS coverage depth problem in \cite{ref5} that addresses the minimum coverage depth in uniform distribution channels under non-random access setting to derive the expectation value of $K_a^{\boldsymbol{p}_t}(n,m)$ in a noiseless channel, as well as its lower bounds in a noisy channel in Section \uppercase\expandafter{\romannumeral4}.

\section{Simulation of DNA Storage Channel Probability Distribution}
According to the definition of the channel probability distribution in Section \uppercase\expandafter{\romannumeral2}, it is closely related to the proportion of each designed strand in the population after PCR amplification. In this section, by analyzing PCR and sequencing data, we find that the real channel probability distribution follows a log-normal distribution. Based on this, we propose a theoretical model for the simulation of the real channel probability distribution, which serves as the foundation for the subsequent research on minimizing sequencing coverage depth.
\subsection{Analysis of Channel Probability Distribution Based on PCR and Sequencing Data}
We first analyze the real channel probability distribution based on the PCR and sequencing data. We synthesized 11,520 oligos, each 150 bases long, using inkjet printing technology at Twist Bioscience (i.e., $n=11520$), and PCR amplification was performed using these oligos as template strands. After 10 cycles of amplification, sequencing 4,970,786 reads to generate Dataset PCR10; after 30 cycles, sequencing 11,001,029 reads to generate Dataset PCR30; and after 60 cycles, sequencing 11,180,177 reads to generate Dataset PCR60. The relationship between different PCR cycles and the copy numbers of different oligos at that cycle is shown in Fig. 2(a), 2(b) and 2(c).

\begin{figure}[htbp]
    \centering
    \subfloat[PCR10]{\includegraphics[width=0.32\linewidth]{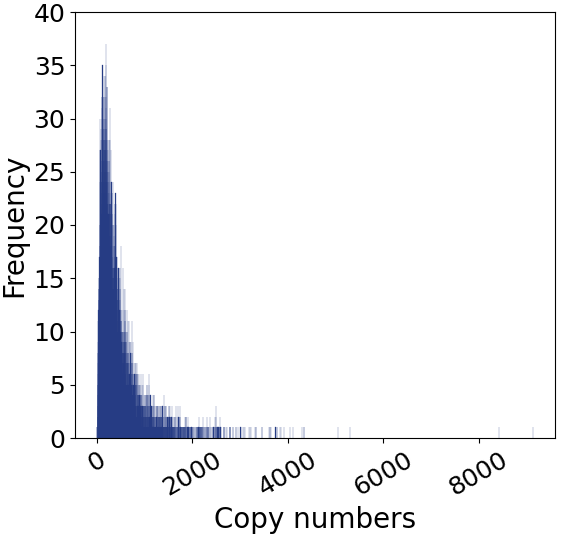}}
    \subfloat[PCR30]{\includegraphics[width=0.32\linewidth]{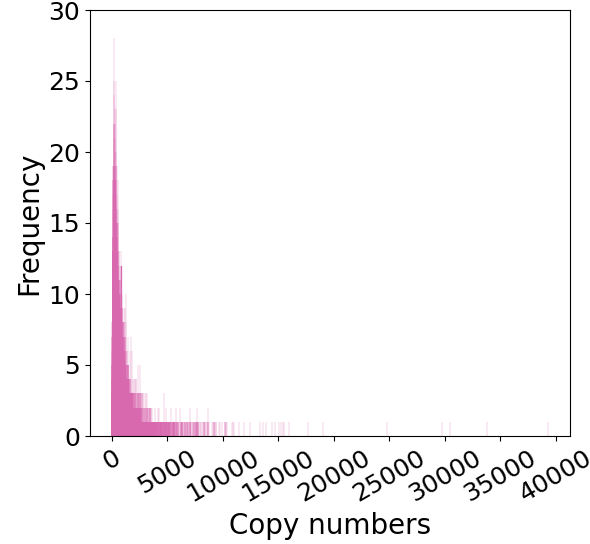}}
    \subfloat[PCR60]{\includegraphics[width=0.32\linewidth]{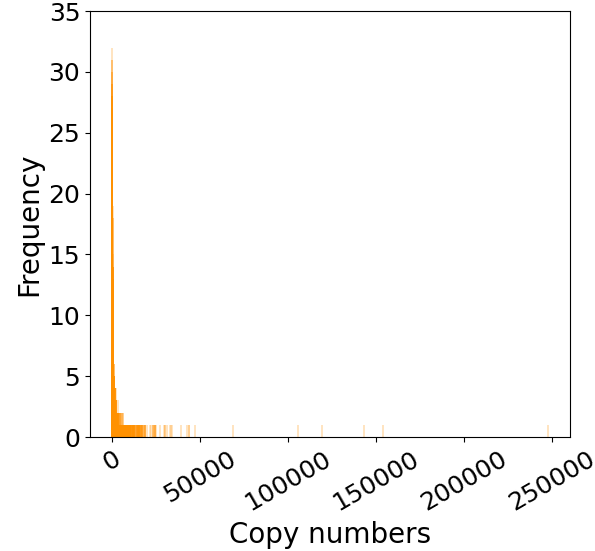}}\\
    \subfloat[Log. PCR10]{\includegraphics[width=0.32\linewidth]{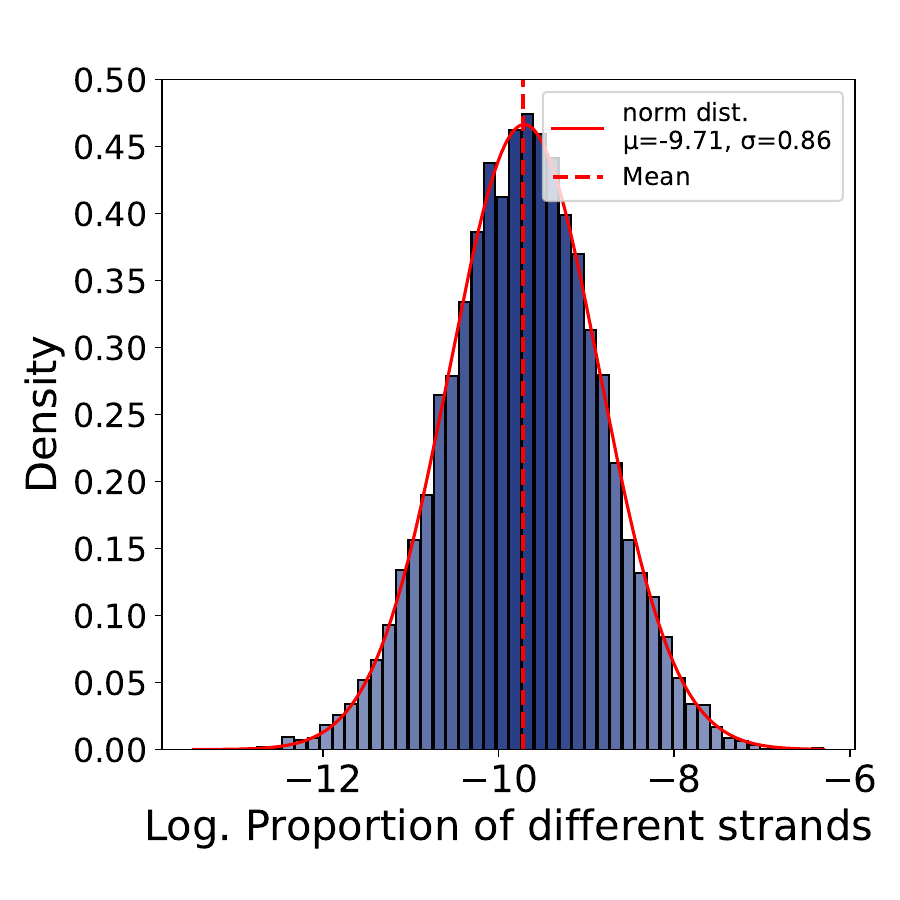}}
    \subfloat[Log. PCR30]{\includegraphics[width=0.32\linewidth]{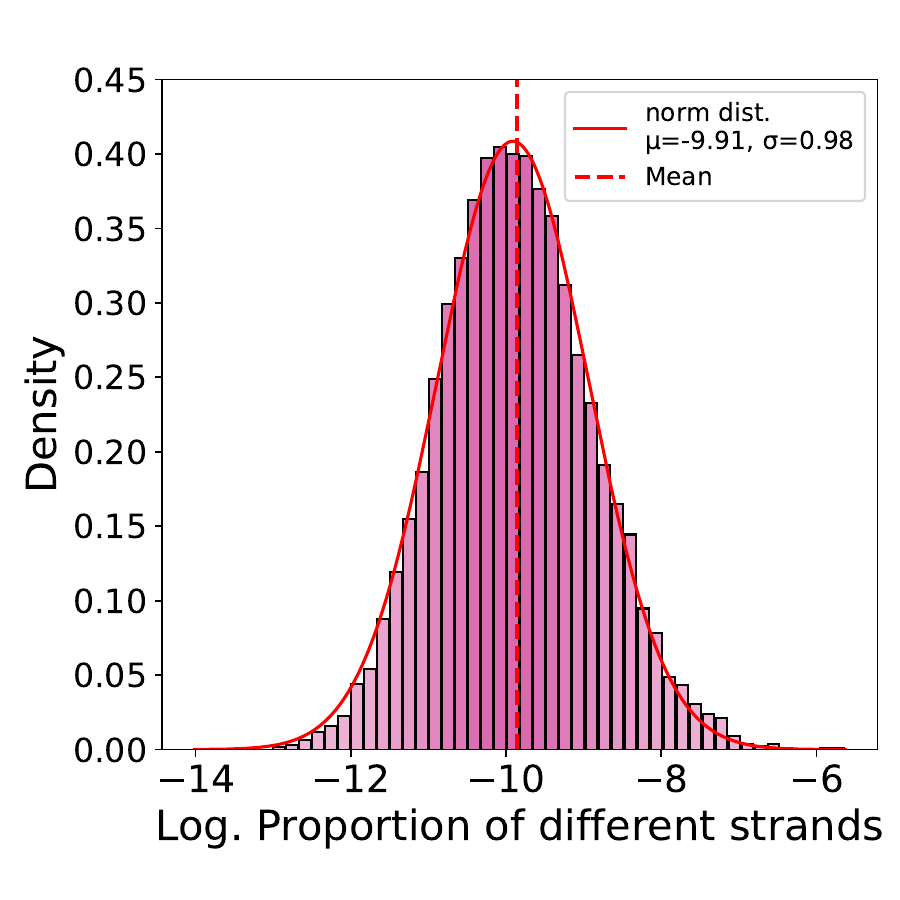}}
    \subfloat[Log. PCR60]{\includegraphics[width=0.32\linewidth]{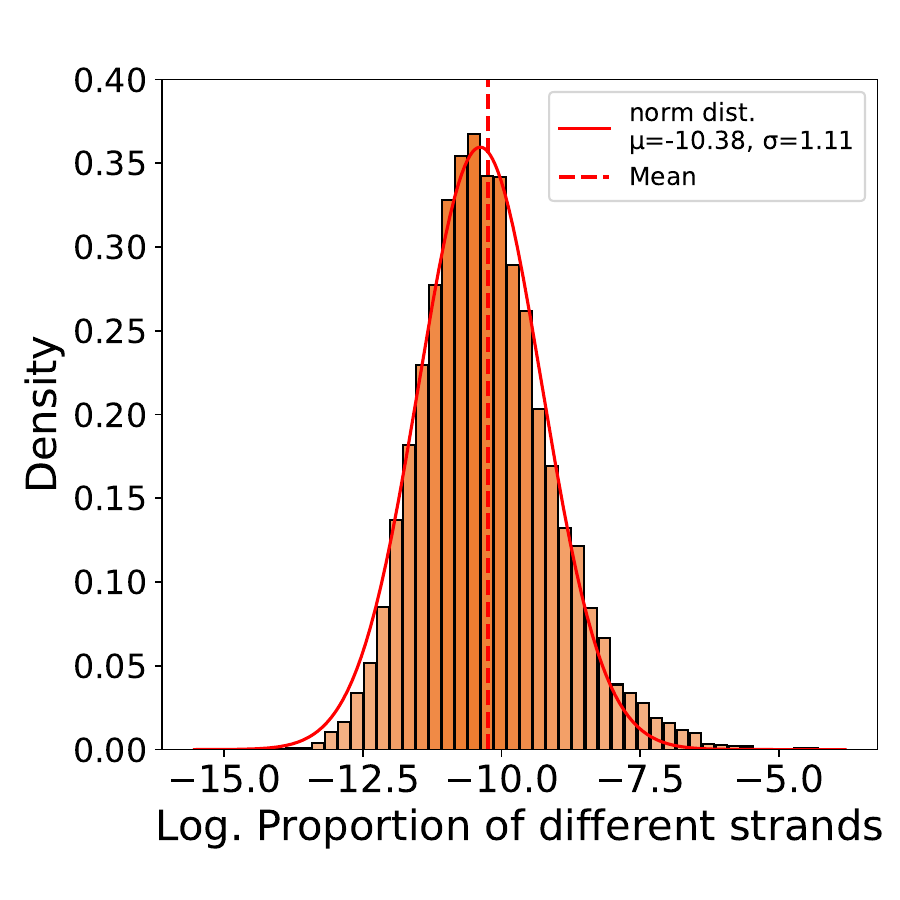}}\\
    \caption{Visualization of different cycles of PCR data.}
    \label{fig:XXX}
\end{figure}

Our primary interest lies in the relationship between the number of PCR cycles and the proportion of each designed strand in the population. Given that the sequencing data exhibit a clear skewed distribution, we assume that the PCR data follow a log-normal distribution \cite{ref16}. Then we performed a logarithmic transformation on the normalized data before fitting a normal distribution, and Fig. 2(d), 2(e) and 2(f) show the results of a good fit on all three datasets, which means after PCR amplification, the channel actually follows a log-normal distribution.

Assuming $X$ is a random variable that follows a normal distribution with parameters $\mu$ and $\sigma^2$ as its expectation and variance after logarithmic transformation, we denote it as $X \sim \mathcal{LN}(\mu,\sigma^2)$. After fitting the PCR data to a log-normal distribution, the sample distributions of the proportion of the $i$-th strand in the population after different PCR cycles are presented in Table \ref{tab:table1}.

Since all three datasets are samples from the overall population generated by PCR amplification for 10, 30, and 60 cycles, we will next perform maximum likelihood estimation (MLE) to estimate the parameters of the population. Let $x_1,x_2,\dots,x_n$ be a simple random sample from the population $X \sim \mathcal{LN}(\mu,\sigma^2)$, the MLE for $\mu$ and $\sigma^2$ are denoted as $\hat{\mu}_{MLE}$ and $\hat{\sigma}^2_{MLE}$ respectively. Then, from \cite{ref17}, we have
\begin{equation}
\label{equ2}
    \hat{\mu}_{MLE}=\frac{1}{n}\sum_{i=1}^n \ln{(x_i)},
\end{equation}
and
\begin{equation}
\label{equ3}
    \hat{\sigma}^2_{MLE}=\frac{1}{n}\sum_{i=1}^n(\ln{(x_i)}-\hat{\mu}_{MLE})^2.
\end{equation}

Based on the calculations from (\ref{equ2}) and (\ref{equ3}), the population distributions of the proportions of different strands after 10, 30, and 60 PCR cycles are presented in Table \ref{tab:table1}.

\begin{table}[H]
    \caption{The distribution of different cycles of PCR data.}
    \label{tab:table1}
    \centering
    \renewcommand{\arraystretch}{1.5} 
    \begin{tabular}{c c c }
        \toprule
        Cycles & Sample distribution & Population distribution \\
        \hline
        PCR10 & $\mathcal{LN}(-9.71,0.86^2)$ & $\mathcal{LN}(-9.72,0.74^2)$ \\
        \hline
        PCR30 & $\mathcal{LN}(-9.91,0.98^2)$ & $\mathcal{LN}(-9.86,0.96^2)$ \\
        \hline
        PCR60 & $\mathcal{LN}(-10.38,1.11^2)$ & $\mathcal{LN}(-10.25,1.38^2)$ \\
        \bottomrule
    \end{tabular}
\end{table}

\subsection{Real Channel Probability Distribution Model}
In the previous subsection, we analyzed PCR experimental data to obtain the real channel distributions after 10, 30, and 60 cycles of PCR amplification. However, in practical experiments, additional sequencing of PCR data to obtain the parameter information of the probability distribution is typically not performed. This necessitates the simulation of the channel probability distribution under the condition that only the synthetic amount $c_i$ and amplification efficiency $r_i$ of each designed strand are known, where PCR amplification efficiency is the rate at which the target DNA fragment is amplified during each cycle in the exponential phase. Theoretically, perfect replication would result in 100\% efficiency, doubling the DNA amount per cycle. However, in practice, efficiency is typically below 100\% and ranges from 80\% to 110\% due to various factors.

Therefore, for given $c_i$ and $r_i$, where $i\in [1,n]$, we model the channel probability distribution as a function of PCR cycles. We first provide the calculation process for the expected copy number of the $i$-th strand after $t$ cycles of PCR amplification, i.e., $\nu_i^{(t)}$.

According to the definition of amplification efficiency, let $\mathbb{E}\left[\nu_i^{(0)}\right]=c_i$, for $t=1$:
\begin{equation*}
    \mathbb{E}\left[\nu_i^{(1)}\right]=\mathbb{E}\left[\nu_i^{(0)}\right] \cdot (1+r_i)=c_i(1+r_i).    
\end{equation*}

For $t=2$:
\begin{equation*}
    \mathbb{E}\left[\nu_i^{(2)}\right]=\mathbb{E}\left[\nu_i^{(1)}\right] \cdot (1+r_i)=c_i(1+r_i)^2.    
\end{equation*}

And so on, for $t=q$:
\begin{equation*}
    \mathbb{E}\left[\nu_i^{(q)}\right]=\mathbb{E}\left[\nu_i^{(q-1)}\right] \cdot (1+r_i)=c_i(1+r_i)^q \quad \text{for all positive integers q}.
\end{equation*}

Thus, we can derive the proportion of the $i$-th strand in the population after $t$ cycles of PCR amplification as
\begin{equation*}
    p_i^{(t)}=\frac{c_i(1+r_i)^t}{\sum_{j=1}^n c_j(1+r_j)^t},    
\end{equation*}
then the expectation and variance of $p_i^{(t)}$ can be calculated, denoted by $\mathbb{E}\left[p_i^{(t)}\right]$ and $\mathrm{Var}\left[p_i^{(t)}\right]$.\\

According to \cite{ref17}, we can calculate the parameters $\mu^{(t)}$ and $\sigma^{(t)}$ of log-normal distribution is that,
\begin{equation}
\label{equ4}
    \mu^{(t)}=\log{\left(\mathbb{E}\left[p_i^{(t)}\right]\right)}-\frac{1}{2}\log{\left(1+\frac{\mathrm{Var}\left[p_i^{(t)}\right]}{\mathbb{E}^2\left[p_i^{(t)}\right]}\right)},
\end{equation}
and
\begin{equation}
\label{equ5}
    \sigma^{(t)}=\sqrt{\ln{\left(1+\frac{\mathrm{Var}\left[p_i^{(t)}\right]}{\mathbb{E}^2\left[p_i^{(t)}\right]}\right)}}.
\end{equation}
Then from the conclusion we obtained in the previous subsection, we have 
\begin{equation*}
    p_i^{(t)} \sim \mathcal{LN}\left(\log{\left(\mathbb{E}\left[p_i^{(t)}\right]\right)}-\frac{1}{2}\log{\left(1+\frac{\mathrm{Var}\left[p_i^{(t)}\right]}{\mathbb{E}^2\left[p_i^{(t)}\right]}\right)},\ln{\left(1+\frac{\mathrm{Var}\left[p_i^{(t)}\right]}{\mathbb{E}^2\left[p_i^{(t)}\right]}\right)}\right).
\end{equation*}
\textbf{Remark}. For the simulation of theorems in the following section, we use the population distribution of PCR amplification and sequencing data from those three different cycles.
\section{MDS Coverage Depth Problem in the Log-normal Distribution Channel}
Under the condition of non-random access, the main goal of our study is to find the minimum sequencing sample size $K$ that ensures the successful decoding of all digital data in the real channel that follows the log-normal distribution after PCR. In this section, we first prove the expected coverage depth in log-normally distributed noiseless channels, and propose the problem of decoding all data successfully in a single sequencing experiment under the expected sample size. Then we prove the theoretical lower bounds of expected sample size in the real noisy channels. For given $n$ designed strands, we denote $\alpha \triangleq \frac{K}{n}$ as the sequencing coverage depth.
\subsection{MDS Coverage Depth Problem in the Noiseless Channel}
In practical experiments, in addition to focusing on the expected sequencing coverage depth required to decode all digital information (i.e., $\mathbb{E}[\alpha]$), we are also concerned with the probability of successfully decoding all data in a single sequencing run (i.e., $\mathrm{Var}[\alpha]$). Let $N$ represent the number of different designed strands when sequencing fixed $K$ reads.\\
\textbf{Remark}. In a noiseless channel, we assume that only one copy of each designed strand is sufficient for successful retrieval of that strand, i.e., $a=1$.\\
\textbf{Theorem 1.} For any channel probability distribution $\boldsymbol{p}_t=(p_1^{(t)},p_2^{(t)},\dots,p_n^{(t)})$ and any $K \ge n \ge m \ge 1$, it holds that
\begin{equation*}
N \sim \mathcal{N} \Bigg(\sum_{i=1}^{n} \left(1 - e^{-K \cdot p_i^{(t)}}\right), \sum_{i=1}^{n} e^{-K \cdot p_i^{(t)}} \left(1 - e^{-K \cdot p_i^{(t)}}\right) - K \left(\sum_{i=1}^{n} p_i^{(t)} e^{-K \cdot p_i^{(t)}}\right)^2 \Bigg).
\end{equation*}
\textit{Proof}. This problem is equivalent to determining the number of urns containing at least one ball after throwing $K$ balls into $n$ labeled urns (one ball per round). Notice that the probability of an identical ball falling into each urn is not equal, and they are given by $\boldsymbol{p}_t=\left(p_1^{(t)},p_2^{(t)},\dots,p_n^{(t)}\right)$, where $p_i^{(t)}$ represents the probability of the ball falls into the $i$-th urn in each round.\\
The value of $\mathbb{E}[N]$: Let
\begin{equation*}
    {X_i^{(K)}}=
    \begin{cases}
        1,&{\text{the $i$-th urn contains at least one ball after $K$ rounds}},\\
        0,&{\text{the $i$-th urn remains empty after $K$ rounds}.}
    \end{cases}
\end{equation*}
thus
\begin{equation*}
    \mathbb{E}[X_i^{(K)}]=1-\left(1-p_i^{(t)}\right)^K \approx 1-e^{K\cdot p_i^{(t)}},
\end{equation*}
when $p_i^{(t)}$ is sufficiently small. Suppose there are $N$ urns containing at least one ball after $K$ rounds, then given $K$,
\begin{equation*}
    \mathbb{E}[N]=\mathbb{E}\left[\sum_{i=1}^n X_i^{(K)}\right]=\sum_{i=1}^n\mathbb{E}[X_i^{(K)}]=\sum_{i=1}^n\left(1-e^{K\cdot p_i^{(t)}}\right).
\end{equation*}
The value of $\mathrm{Var}[N]$: Inspired by \cite{ref18}, we first investigate the number of urns, denote it as $N_0$, that remain empty after throwing $K$ balls into $n$ urns. According to \cite{ref18},
\begin{equation*}
    \mathrm{Var}[N_0]=\sum_{i=1}^n \left(e^{-K p_i^{(t)}}-e^{-2K p_i^{(t)}}\right)-K\left(\sum_{i=1}^n p_i^{(t)}e^{-K p_i^{(t)}}\right)^2.
\end{equation*}
Thus
\begin{equation*}
    \mathrm{Var}[N]=\mathrm{Var}[n-N_0]=\mathrm{Var}[N_0]= \sum_{i=1}^n \left(e^{-K p_i^{(t)}}-e^{-2K p_i^{(t)}}\right)-K\left(\sum_{i=1}^n p_i^{(t)}e^{-K p_i^{(t)}}\right)^2.
\end{equation*}
By the central limit theorem,
\begin{equation*}
    N\sim \mathcal{N}\left(\sum_{i=1}^n\left(1-e^{K\cdot p_i^{(t)}}\right),\sum_{i=1}^n \left(e^{-K p_i^{(t)}}-e^{-2K p_i^{(t)}}\right)-K\left(\sum_{i=1}^n p_i^{(t)}e^{-K p_i^{(t)}}\right)^2\right).
\end{equation*}
$\hfill\blacksquare$

On the basis of Theorem 1, let $\boldsymbol{p}_t=\left(p_1^{(t)},p_2^{(t)},\dots,p_n^{(t)}\right)=\left(\frac{1}{n},\frac{1}{n},\dots,\frac{1}{n}\right)$, the following result is straightforward.\\
\textbf{Theorem 2.} For a uniform distribution channel, and any $K \ge n \ge m \ge 1$, it holds that
\begin{align*}
    N\sim \left(n\left(1-e^{-\alpha}\right),n(e^{-\alpha}-e^{-2\alpha}-\alpha e^{-2\alpha})\right).
\end{align*}

In summary, for $n$ designed strands, when sequencing fixed $K$ reads, the expected ratio of successfully decoding designed strands is $\frac{\mathbb{E}[N]}{n}$. That is, if $m$ information strands are encoded with the $\left[\frac{n \cdot m}{\mathbb{E}[N]},m\right]$ MDS code, then theoretically speaking, sequencing $K$ reads in a noiseless channel recovers complete information.

Applying the conclusions of Theorem 1 and Theorem 2, we present in Fig. 3 a comparison of the minimum sequencing coverage depth under real channels and a uniform distribution channel. Fig. 4 visualizes the variance derivative function to intuitively show the probability of successfully decoding all digital data in a single sequencing experiment.

Based on the applications of Theorem 1 and Theorem 2, we obtain the following conclusions:
\begin{enumerate}
    \item When encoding $m$ information strands with identical redundancy, the minimum sequencing coverage depth required to decode all data in a uniform distribution channel is considerably smaller than in a real channel. Therefore, further investigation of the minimum sequencing coverage depth in a log-normal distribution channel is essential.
    \item The higher the encoding redundancy, the smaller the minimum sequencing coverage depth in any channels.
    \item In a log-normal distribution channel, the minimum sequencing coverage depth increases with the number of PCR cycles. This is attributed to varying amplification efficiencies among the designed strands, leading to the over-amplification and sequencing of a small subset of strands as the cycle count rises.
    \item The variance reaches its maximum between $\alpha =1$ and $\alpha =2$ for PCR cycles between 10 and 60, and tends to minimize when $\alpha >7$ and $\alpha >8$ respectively, which means that when the expected coverage depth is between 1 and 2, although sequencing $\alpha m$ reads is sufficient to decode all the data, the probability of successfully decoding all the data in a single experiment is minimized. We recommend increasing the sequencing ratio or decreasing the encoding redundancy to reduce the variance.
\end{enumerate}
\begin{figure}[htbp]
    \centering
    \includegraphics[width=0.75\linewidth]{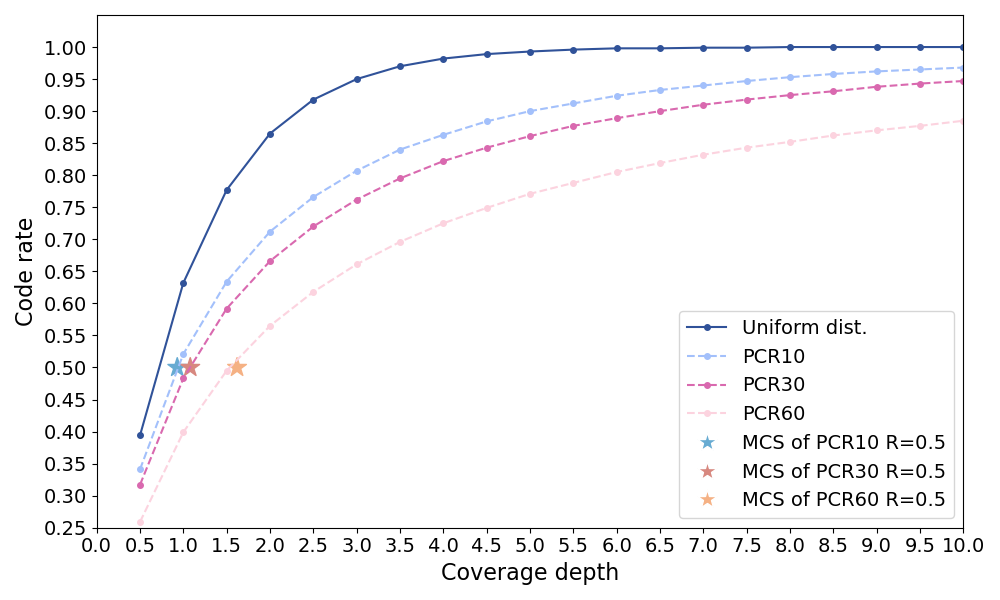}
    \caption{Expected coverage depth required for successful decoding under different code rates. The \textit{Uniform dist.} curve represents the relationship between coverage depth and code rate in a uniform distribution channel. The \textit{PCR10}, \textit{PCR30}, and \textit{PCR60} curves correspond to the empirical channels derived from Dataset PCR10, Dataset PCR30, and Dataset PCR60, respectively, showing how coverage depth varies with code rate. Three points denote Monte Carlo simulations performed under the respective channel with code rate $R = 0.5$.}
    \label{fig:expectation}
\end{figure}
\begin{figure}[htbp]
    \centering
    \includegraphics[width=0.75\linewidth]{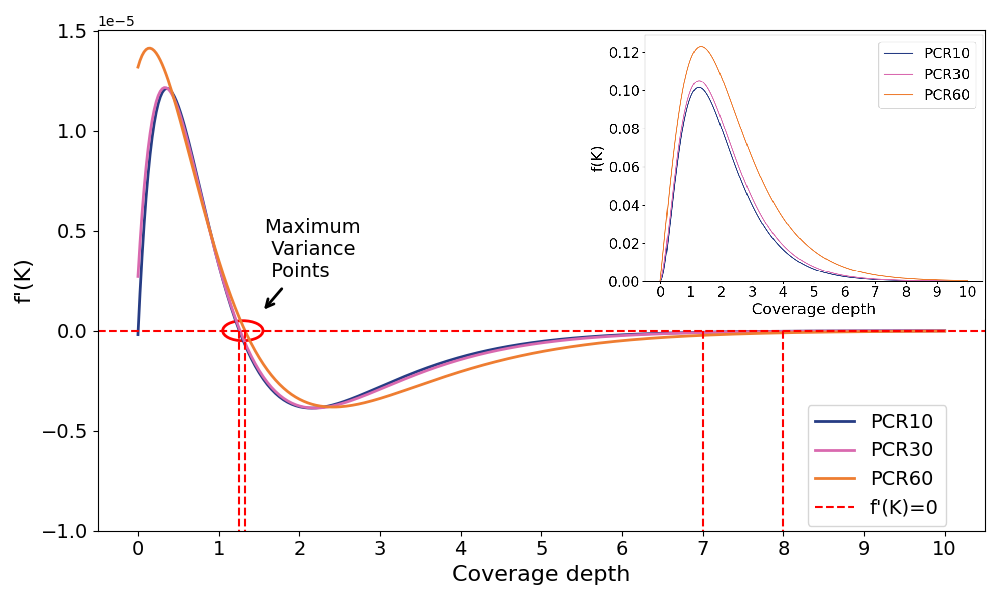}
    \caption{The inner plot illustrates the probability that a single experiment fails to decode all information when sequencing is performed at the expected sample size. This probability is quantified by the variance of each designed strand under that sequencing sample size, denoted as $f(K) \triangleq e^{-K \mathbb{E}\left[p_i^{(t)}\right]}-e^{-2K \mathbb{E}\left[p_i^{(t)}\right]}-\frac{n}{K}\left(K\mathbb{E}\left[p_i^{(t)}\right]\right)^2e^{-2K \mathbb{E}\left[p_i^{(t)}\right]}$. The outer plot depicts the trend of this probability, i.e., the derivative $f'(K)$. The outer plot indicates that the variance has only one peak, corresponding to the maximum variance point as annotated in the figure.}
    \label{fig:Var}
\end{figure}

\subsection{MDS Coverage Depth Problem in the Noisy Channel}
Theorem 3 and Theorem 4 of this section present two lower bounds on $K_a^{\boldsymbol{p}_t}(n,m)$ in noisy channels obeying a log-normal distribution.\\
\noindent \textbf{Remark}. In a noisy channel, we assume that at least $a$ noisy copies of each designed strand need to be sequenced for successful retrieval of that strand, i.e., $a > 1$.

According to Section \uppercase\expandafter{\romannumeral3}, if given the synthetic amount and the amplification efficiency of each strand, denoted as $\boldsymbol{c}=(c_1,c_2,\dots,c_n)$,  $\boldsymbol{r}=(r_1,r_2,\dots,r_n)$ respectively, we can deduce the parameters $\mu^{(t)}$ and $\sigma^{(t)}$ of the real log-normal distribution channel after $t$ cycles of PCR.

From \cite{ref19}, we know that if $p_i^{(t)} \sim \mathcal{LN}\left(\mu^{(t)}, {\sigma^{(t)}}^2\right)$, then the expected value of $p_i^{(t)}$ is
\begin{equation*}
    \mathbb{E}\left[p_i^{(t)}\right]=e^{\mu^{(t)} + \frac{{\sigma^{(t)}} ^2}{2}},
\end{equation*}
and the expected value of $\frac{1}{p_i^{(t)}}$ is
\begin{equation*}
    \mathbb{E}\left[\frac{1}{p_i^{(t)}}\right]=e^{-\mu^{(t)} + \frac{{\sigma^{(t)}} ^2}{2}}.
\end{equation*}
\noindent \textbf{Theorem 3}. For given $\boldsymbol{c}$, $\boldsymbol{r}$, for any $t\ge 1$, $\beta>1$, $a>1$, let $R=\frac{m}{n}$, it holds that $P\left[K_a^{\boldsymbol{p}_t}(n,m) \le K_1\left(\mu^{(t)},\sigma^{(t)},a,R\right)\right] \le e^{-\beta}\left(1+\frac{m}{n-m}\right)$, where
\begin{equation*}
    K_1(\mu^{(t)},\sigma^{(t)},a,R)\triangleq e^{-\mu^{(t)} +\frac{{\sigma^{(t)}}^2}{2}}\left(\log\Bigg(\frac{1}{1-R}\right)-\beta\Bigg).
\end{equation*}
\textit{Proof}. Since $K_1^{\boldsymbol{p}_t}(n,m)<K_a^{\boldsymbol{p}_t}(n,m)$, the following proves that
\begin{equation*}
    P\left[K_1^{\boldsymbol{p}_t}(n,m) \le K_1\right] \le e^{-\beta}\left(1+\frac{m}{n-m}\right).
\end{equation*}
Inspired by \cite{ref5}, we have that,
\begin{align*}
    e^{\log\left(\frac{n}{n-m}\right)-\beta} \cdot \mathbb{E}\left[e^{-\mathbb{E}\left[p_i^{(t)}\right]\cdot K_1^{\boldsymbol{p}_t}(n,m)}\right]
    &=e^{\log\left(\frac{n}{n-m}\right)-\beta} \cdot \sum_{k=1}^{\infty} e^{-k \cdot \mathbb{E}\left[p_i^{(t)}\right]}P\left[K_1^{\boldsymbol{p}_t}(n,m)=k\right]\\
    & = \sum_{k=1}^{\infty} e^{\log\left(\frac{n}{n-m}\right)-\beta-k \cdot \mathbb{E}\left[p_i^{(t)}\right]}P\left[K_1^{\boldsymbol{p}_t}(n,m)=k\right]\\
&= \sum_{k=1}^{\left \lfloor \mathbb{E}\left[\frac{1}{p_i^{(t)}}\right] \cdot \log\left(\frac{n}{n-m} \right) - \mathbb{E}\left[\frac{1}{p_i^{(t)}}\right]\cdot \beta \right \rfloor}e^{\log\left(\frac{n}{n-m}\right)-\beta-k \cdot \mathbb{E}\left[p_i^{(t)}\right]}P\left[K_1^{\boldsymbol{p}_t}(n,m)=k\right]\\
    &+ \sum_{k=1+\left \lfloor \mathbb{E}\left[\frac{1}{p_i^{(t)}}\right] \cdot \log\left(\frac{n}{n-m} \right) - \mathbb{E}\left[\frac{1}{p_i^{(t)}}\right]\cdot \beta \right \rfloor}^{\infty}e^{\log\left(\frac{n}{n-m}\right)-\beta-k \cdot \mathbb{E}\left[p_i^{(t)}\right]}P\left[K_1^{\boldsymbol{p}_t}(n,m)=k\right]\\ 
    &\ge \sum_{k=1}^{\left \lfloor \mathbb{E}\left[\frac{1}{p_i^{(t)}}\right] \cdot \log\left(\frac{n}{n-m} \right) - \mathbb{E}\left[\frac{1}{p_i^{(t)}}\right]\cdot \beta \right \rfloor}e^{\log\left(\frac{n}{n-m}\right)-\beta-k \cdot \mathbb{E}\left[p_i^{(t)}\right]}P\left[K_1^{\boldsymbol{p}_t}(n,m)=k\right]\\
    &\ge \sum_{k=1}^{\left \lfloor \mathbb{E}\left[\frac{1}{p_i^{(t)}}\right] \cdot \log\left(\frac{n}{n-m} \right) - \mathbb{E}\left[\frac{1}{p_i^{(t)}}\right]\cdot \beta \right \rfloor}1 \cdot P\left[K_1^{\boldsymbol{p}_t}(n,m)=k\right]\\
    & \ge P\left[K_1^{\boldsymbol{p}_t}(n,m) \le \mathbb{E}\left[\frac{1}{p_i^{(t)}}\right] \cdot \log\left(\frac{n}{n-m} \right) - \mathbb{E}\left[\frac{1}{p_i^{(t)}}\right]\cdot \beta\right]\\
    &=P\left[K_1^{\boldsymbol{p}_t}(n,m) \le e^{-\mu^{(t)} + \frac{{\sigma^{(t)}} ^2}{2}} \cdot \log\left(\frac{n}{n-m} \right) - e^{-\mu^{(t)} + \frac{{\sigma^{(t)}} ^2}{2}}\cdot \beta\right].
\end{align*}
From \cite{ref20}, the generating function of the geometric random variable $K_1^{\boldsymbol{p}_t}(n,m)$ is given by
\begin{align*}
    G_{K_1^{\boldsymbol{p}_t}(n,m)}\left(e^{-\mathbb{E}\left[p_i^{(t)}\right]}\right) &= \mathbb{E}\left[{e^{-\mathbb{E}\left[p_i^{(t)}\right]\cdot{K_1^{\boldsymbol{p}_t}(n,m)}}}\right]\\
    &=\sum_{k=0}^{\infty} P\left[K_1^{\boldsymbol{p}_t}(n,m)=k\right]{\left(e^{-\mathbb{E}\left[p_i^{(t)}\right]}\right)}^k\\
    & =  \prod_{j=1}^m \frac{1-\frac{j-1}{n}}{e^{\mathbb{E}\left[p_i^{(t)}\right]}-\frac{j-1}{n}} \\
    &\le \prod_{j=1}^m \frac{1-\frac{j-1}{n}}{1+\mathbb{E}\left[p_i^{(t)}\right]-\frac{j-1}{n}} \le 1,
\end{align*}
where the inequality is obtained from Taylor expansion of $e^{\mathbb{E}\left[p_i^{(t)}\right]}$. Finally, we conclude that,
\begin{align*}
    P\left[K_a^{\boldsymbol{p}_t}(n,m) \le K_1\left(\mu^{(t)},\sigma^{(t)},a,R\right)\right]&\le e^{\log\left(\frac{n}{n-m}\right)-\beta} \cdot \mathbb{E}\left[e^{-\mathbb{E}\left[p_i^{(t)}\right]\cdot K_1^{\boldsymbol{p}_t}(n,m)}\right]\\
    & \le e^{-\beta} \cdot \frac{n}{n-m} \cdot 1 \le e^{-\beta}\left(1+\frac{m}{n-m}\right).
\end{align*}$\hfill \blacksquare$

Let $N_K$ represent the number of urns that contain less than $a$ balls after $K$ rounds. When using an $[n,m]$ MDS code, recovering all data means successfully decoding $m$ out of $n$ strands, implying that at most $n-m$ strands cannot be successfully decoded.\\
\textbf{Theorem 4}. For given $\boldsymbol{c}$, $\boldsymbol{r}$, for any $a>1$, $t\ge 1$, let $R=\frac{m}{n}$, we have $\mathbb{E}[N_K] \le n-m$, if $K_a^{\boldsymbol{p}_t}(n,m) > K_2\left(\mu^{(t)},\sigma^{(t)},a,R\right)$, where
\begin{equation*}
    K_2\left(\mu^{(t)},\sigma^{(t)},a,R\right) \triangleq e^{-\mu^{(t)} +\frac{{\sigma^{(t)}}^2}{2}}\Bigg((a-1)-\log2\log(1-R)+(a-1)\sqrt{-\frac{2\log 2}{a-1}\log(1-R)}\Bigg).
\end{equation*}

Before proving Theorem 4, we will first present a lemma, which will be used in the proof of Theorem 4.\\
\textbf{Lemma 1}. For $K > (a-1) \cdot \mathbb{E}\left[\frac{1}{p_i^{(t)}}\right]$, we have $\mathbb{E}\left[N_K\right] \le n-m$, if $\frac{K \cdot \mathbb{E}\left[p_i^{(t)}\right]}{a-1}e^{-\frac{K \cdot \mathbb{E}\left[p_i^{(t)}\right]}{a-1}} \le \frac{1}{e}\left(1-R\right)^{\frac{\log2}{a-1}}$.\\
\textit{Proof}. For clarity in writing, we omit the superscript of $p_i^{(t)}$ in the proof of Lemma 1 and denote it as $p_i$. We consider this question in the context of the urn problem. According to Claim 2 in \cite{ref5}, the probability of the $i$-th urn containing less than $a$ balls after $K$ rounds is
\begin{equation*}
    q_i=\sum_{j=0}^{a-1} \binom{K}{j}{p_i}^j(1-p_i)^{K-j} \le e^{-KD\left(\frac{a-1}{K}\Big|\Big|p_i\right)},
\end{equation*}
where
\begin{equation*}
    D(a||b) \triangleq a\log_2 \frac{a}{b}+(1-a)\log_2 \frac{1-a}{1-b}
\end{equation*}
is the Kullback-Leibler divergence \cite{ref21}. Then
\begin{equation*}
    \mathbb{E}[N_K]=\sum_{i=1}^n q_i \le \sum_{i=1}^n e^{-(a-1)\log_2 \frac{a-1}{Kp_i}-(K-(a-1))\log_2 \frac{K-(a-1)}{K\left(1-p_i\right)}} \le ne^{-(a-1)\log_2 \frac{a-1}{K\mathbb{E}\left[p_i\right]}-(K-(a-1))\log_2 \frac{K-(a-1)}{K\left(1-\mathbb{E}\left[p_i\right]\right)}},
\end{equation*}
let $f(p_i)=e^{-(a-1)\log_2 \frac{a-1}{Kp_i}-(K-(a-1))\log_2 \frac{K-(a-1)}{K\left(1-p_i\right)}}$, since $f(p)$ is a convex function, $e^{-f(p)}$ is a log-concave function, therefore, the second inequality follows from Jensen's inequality.

Hence, $\mathbb{E}[N_K] \le n-m=n(1-R)$ holds if and only if
\begin{equation*}
    ne^{-(a-1)\log_2 \frac{a-1}{K\mathbb{E}\left[p_i\right]}-(K-(a-1))\log_2 \frac{K-(a-1)}{K\left(1-\mathbb{E}\left[p_i\right]\right)}}\le n(1-R),
\end{equation*}
that is,
\begin{equation}
\label{equ12}
    e^{-(a-1)\log_2 \frac{a-1}{K\mathbb{E}\left[p_i\right]}-(K-(a-1))\log_2 \frac{K-(a-1)}{K\left(1-\mathbb{E}\left[p_i\right]\right)}}\le 1-R.
\end{equation}

By the change of base formula, the following equality holds.
\begin{equation*}
    e^{-(a-1)\log_2 \frac{a-1}{K\mathbb{E}\left[p_i\right]}-(K-(a-1))\log_2 \frac{K-(a-1)}{K\left(1-\mathbb{E}\left[p_i\right]\right)}}=\left(\frac{K\mathbb{E}\left[p_i\right]}{a-1}\right)^{\frac{a-1}{\ln{2}}}\left(\frac{K \cdot \frac{1}{\mathbb{E}\left[p_i\right]}-K}{K \cdot \frac{1}{\mathbb{E}\left[p_i\right]}-(a-1)\cdot \frac{1}{\mathbb{E}\left[p_i\right]}}\right)^{\frac{K-(a-1)}{\ln{2}}}.
\end{equation*}
Let $K \triangleq \lambda(a-1)\cdot \frac{1}{\mathbb{E}\left[p_i\right]}$, then (\ref{equ12}) can be rewrite as
\begin{equation}
\label{equ13}
    \lambda \left(1-\frac{\lambda-1}{\lambda\cdot \frac{1}{\mathbb{E}\left[p_i\right]}-1}\right)^{\lambda\cdot \frac{1}{\mathbb{E}\left[p_i\right]}-1} \le (1-R)^{\frac{\ln{2}}{a-1}}.
\end{equation}
According to the definition of e, we know that
\begin{equation*}
   \left(1-\frac{\lambda-1}{\lambda\cdot \frac{1}{\mathbb{E}\left[p_i\right]}-1}\right)^{\lambda\cdot \frac{1}{\mathbb{E}\left[p_i\right]}-1} \le e^{-(\lambda-1)}.
\end{equation*}
Thus, if $\lambda e^{-\lambda} \le \frac{1}{e}(1-R)^{\frac{\ln{2}}{a-1}}$ holds then (\ref{equ13}) also holds.

By the assumption,
\begin{equation*}
    \frac{K\mathbb{E}\left[p_i\right]}{a-1}e^{\frac{K\mathbb{E}\left[p_i\right]}{a-1}} \le \frac{1}{e}(1-R)^{\frac{\ln{2}}{a-1}}.
\end{equation*}$\hfill \blacksquare$

Based on Lemma 1, we will prove Theorem 4 next.\\
\textit{Proof of Theorem 4}. Let $x=\frac{1}{e}\left(1-R\right)^{\frac{\log2}{a-1}}$, $y=\frac{K\cdot \mathbb{E}\left[p_i^{(t)}\right]}{a-1}$, according to Claim 7 in \cite{ref5}, the equation
\begin{equation*}
    -\frac{K \cdot \mathbb{E}\left[p_i^{(t)}\right]}{a-1}e^{-\frac{K \cdot \mathbb{E}\left[p_i^{(t)}\right]}{a-1}}=-ye^{-y}=-x=-\frac{1}{e}\left(1-R\right)^{\frac{\log2}{a-1}}
\end{equation*}
has two solutions $y=-W_0(-x)$ and $y=-W_{-1}(-x)$. From Claim 8 in \cite{ref5}, we have that,
\begin{equation*}
    -W_{-1}(-x)<1+\sqrt{-\frac{2\log2}{a-1}\log\left(1-R\right)}-\frac{\log2}{a-1}\log\left(1-R\right).
\end{equation*}
The function $-ye^{-y}$ is both continuous and monotonically increasing for $y$. Therefore, for any given $R=\frac{m}{n}$, only the branch $W_{-1}$ is applicable. This means that for $K \ge (a-1) \cdot \mathbb{E}\left[\frac{1}{p_i^{(t)}}\right]$, Lemma 1 holds if and only if $y \ge -W_{-1}(-x)$. Hence, according to the expression of $y$, a sufficient condition for $\mathbb{E}\left[N_K\right] \le n-m$ is
\begin{equation*}
    \frac{K\cdot \mathbb{E}\left[p_i^{(t)}\right]}{a-1} \ge 1+\sqrt{-\frac{2\log2}{a-1}\log\left(1-R\right)}-\frac{\log2}{a-1}\log\left(1-R\right),
\end{equation*}
equals to
\begin{equation*}
    K \ge \frac{1}{\mathbb{E}\left[p_i^{(t)}\right]} \cdot (a-1)-\frac{1}{\mathbb{E}\left[p_i^{(t)}\right]} \cdot \log2 \log\left(1-R\right) +\frac{1}{\mathbb{E}\left[p_i^{(t)}\right]} \cdot (a-1)\sqrt{-\frac{2\log2}{a-1}\log\left(1-R\right)}.
\end{equation*}
According to Jensen's inequality, $\frac{1}{\mathbb{E}\left[p_i^{(t)}\right]} \le \mathbb{E}\left[\frac{1}{p_i^{(t)}}\right]$, thus a more sufficient condition for $\mathbb{E}\left[N_K\right] \le n-m$ is
\begin{equation*}
    K \ge \mathbb{E}\left[\frac{1}{p_i^{(t)}}\right] \cdot (a-1)-\mathbb{E}\left[\frac{1}{p_i^{(t)}}\right] \cdot \log2 \log\left(1-R\right) +\mathbb{E}\left[\frac{1}{p_i^{(t)}}\right] \cdot (a-1)\sqrt{-\frac{2\log2}{a-1}\log\left(1-R\right)},
\end{equation*}
that is,
\begin{equation*}
    e^{-\mu^{(t)} +\frac{{\sigma^{(t)}}^2}{2}}\left((a-1)-\log2\log(1-R)+(a-1)\sqrt{-\frac{2\log 2}{a-1}\log(1-R)}\right).
\end{equation*}
$\hfill\blacksquare$

We compare the two lower bounds of the expected sequencing coverage depth that we prove (i.e., $K_1$ and $K_2$ respectively) with the maximum sequencing coverage depth in the real channel obtained through 1000 Monte Carlo simulations (i.e., sim-max) under the condition that $a=2$, $R=0.8$, as shown in Table \ref{tab:table2} below. 

\begin{table}[H]
    \caption{Simulation and theoretical bounds ($a=2$, $R=0.8$).}
    \label{tab:table2}
    \centering
    \renewcommand{\arraystretch}{1.4} 
    \begin{tabular}{c c c c}
        \toprule
        Cycles & $K_1$ & $K_2$ & sim-max \\
        \hline
        PCR10 & 3.33 & 7.51 & 6.12 \\
        \hline
        PCR30 & 4.32 & 9.75 & 7.68 \\
        \hline
        PCR60 & 7.8 & 17.61 & 12.85 \\
        \bottomrule
    \end{tabular}
\end{table}

\section{Conclusion}
In this paper, by working with PCR and sequencing data, we analyze and simulate the probability distribution of the real channel, and mainly investigate the sequencing coverage depth problem based on experimental data, under the non-random access setting. That is, we prove the expected coverage depth and its theoretical lower bounds in real noiseless and noisy channels respectively, and first propose the problem of decoding all data successfully in a single sequencing experiment using the expected coverage depth as the sample size.

\end{document}